\documentclass[aps,prb,showpacs,twocolumn]{revtex4}

\usepackage{graphicx}

\usepackage{amsmath} 
\newcommand{\EqLabel}[1]{\label{#1}}

\begin{document}
\title{Ferromagnetic spin-polaron on  complex lattices}

\author{Mona Berciu and George A. Sawatzky} 

\affiliation{ Department of Physics and Astronomy, University of
  British Columbia, Vancouver, BC, Canada, V6T~1Z1}

\begin{abstract}
We present a simpler derivation of the exact solution of a spin-polaron
in a ferromagnet and generalize it to complex lattices and/or longer
range exchange interactions. As a specific example, we analyze a
two-dimensional MnO$_2$-like lattice (as in the ferromagnetic layers
in LaMnO$_3$) and discuss the properties of the resulting spin-polaron
in various regimes. At strong couplings the solution is reminiscent of
the Zhang-Rice singlet, however the electronic wavefunction involved
in the singlet is dependent on the momentum of the singlet, and
multiple bands may appear.
\end{abstract}

\date{\today} 

\maketitle

\section{Introduction}

The motion of a charged particle in a magnetically ordered background
is one of the central basic problems encountered in doped,
magnetically ordered insulators. Well known examples are hole or
electron doped ferromagnetic insulators such as EuO,\cite{torrance}
hole doped parent 
compounds of the colossal magneto-resistance materials,\cite{cheong}
hole and electron doped parent compounds of high-temperature
superconductors,\cite{mueller,anderson} etc.

The exact solution of the
problem in  a two-dimensional (2D) antiferromagnetic lattice as
represented by the cuprates still eludes us. Diagrammatic Quantum
Monte Carlo calculations on an assumed Neel-ordered lattice and in a
single band $tJ$-like model provide an exact numerical solution to
this approximated system.\cite{prokofev} In the real system, however,
the hole propagates in an O ${2p}$ band while the spins are a result
of a half-filled Cu $d_{x^2-y^2}$ band with a large Hubbard $U$.

The other examples like EuO or LaMnO$_3$ are either ferromagnetic or,
as in the case of  LaMnO$_3$, have ferromagnetic 2D layers. As we show
here, for these
systems there is an exact solution available for a hole or electron
propagating in either the same, or a different band from that of the
spin background, and for any sign and magnitude of the coupling. Exact
solutions of this kind provide important information on the existence
range of bound spin-polaron states as envisioned for example
for Zhang-Rice (ZR) singlets,\cite{ZR} in which the hole propagates in
an O sublattice and the local spins are on a transition metal (TM)
sublattice. It is important to note that aside from cuprates, there is
evidence that doped holes in  manganites LaMnO$_3$ and
cobaltates  Na$_x$CoO$_2$ also propagate  on the O sublattice but are
strongly coupled to the TM. 

Indeed, it has been known for a long time \cite{ra, rb,r1,r2,r3,r4} that an
exact solution can be found for the Green's function of a particle
(electron or hole)
moving in a lattice of ferromagnetically (FM) ordered spins, at zero
temperature. If the spin of the particle is parallel to the FM order,
the solution is trivial: its  energy  is simply shifted
by its exchange coupling to the FM spins. However, if the particle
spin is antiparallel to the FM order, it can scatter and spin-flip by
creating a magnon. Depending on the values of the various parameters
and the total energy, this can result in a finite lifetime (incoherent
scattering leading to a broad spectral weight) or in an
infinitely-lived quasiparticle -- the spin-polaron -- comprising the
bound-state of the particle and the magnon. In the later case,
the dispersion of the spin-polaron can be significantly renormalized
compared to that of the free particle.

Similar physics is found in a seemingly more complex Kondo-Anderson
model that also admits an exact solution.\cite{Ovh} This is not
surprising, because the restriction placed there on the allowed
occupation numbers of the $d$ orbitals essentially maps it back to an
electron interacting with FM-ordered spins.  Effort has also been
focused on trying to extend this type of solution to
finite-temperature ({\it i.e.} presence of multiple magnons), to
finite particle concentrations, etc. While, to our knowledge, no other
exact solutions have been found, such work has resulted in various
approximations for the self-energy.\cite{r5}

One common aspect of all these exact solutions are the assumptions (1) that
the particle moves on the same lattice that hosts the FM spins
(usually a simple cubic-like lattice in $d$ dimensions, although
generalizations to other cases are trivial), and (2) that the
particle-spin exchange is local, {\em i.e.} purely on-site.

In this article we show that the $T=0$ exact solution can be
generalized to systems where these restrictions are lifted. In other
words, to cases where the particle moves on the same lattice that
hosts the FM spins but the exchange is longer range, as well as to
cases where the particle moves on a different sublattice than the one
hosting the FM spins so that the exchange is necessarily not
on-site. We focus on a specific problem of the later type, and briefly 
comment on other possible generalizations later on. While our
formalism  is similar in spirit to that used in
Refs. \onlinecite{ra,rb,r1,r2,r3,r4}, it is in practice much simpler
to use and more transparent. This is essential to allow us to find
these generalized exact solutions.

Such calculations are necessary in order to understand quantitatively
direct/inverse angle-resolved photo-emission spectroscopy (ARPES) (when
the particle is a hole/electron), in insulators which order
ferromagnetically, for instance oxides like EuO, CuCr$_2$S$_4$,
CuCr$_2$Se$_4$, MgFe$_2$O$_4$, etc. They may also be relevant to some
extent for itinerant ferromagnets, given that spin-polarized electron
loss spectroscopy on thin Co films\cite{Krs} reveals good fits of the
measured spin-wave spectrum to effective Heisenberg models, however
more work needs to be done to understand how to properly extend it in
this direction. 

Our results suggest possible spintronic uses for these materials, in
terms of transport of spin-polarized currents, since we prove
that charge carriers with spin-polarization antiparallel to that of
the FM background can propagate {\em coherently}, in other words
scattering on magnons is not necessarily leading to a finite lifetime.
Equally importantly, the results also give some useful insights on ways to
improve our understanding of the propagation of electrons or holes in
antiferromagnetic backgrounds. These issues are discussed in more
detail below.

The paper is organized as follows: in section II we introduce the
specific model to be solved. In section III we give its exact
solution, and comment on various possible generalizations. In
section IV we present and analyze a selection of interesting
results. Section V contains our summary and conclusions.

\section{Model}

\begin{figure}[t]
\includegraphics[width=0.85\columnwidth]{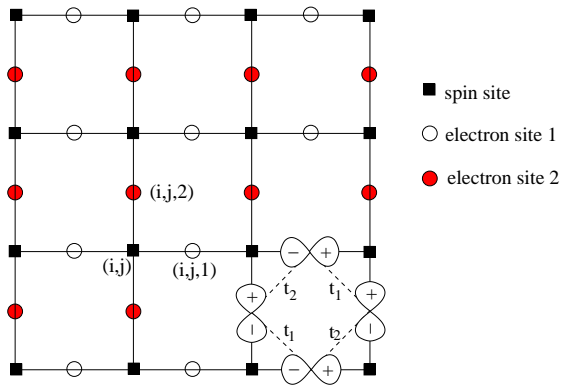}
\caption{Sketch of the lattice and the indexes of
  sites in the unit cell. The bottom-right cell shows
  the signs of the $p$ orbital lobes, which define the signs of the
  hopping integrals $t_1, t_2$. }
\label{fig1}
\end{figure}

We study the 2D model of
Fig. \ref{fig1}. It consists of two sublattices: the one hosting the
FM-ordered spins is a simple square lattice of lattice constant $a$,
whose sites are marked by black squares (the analogs of the Mn sites
in a MnO$_2$ lattice). The charge carrier is moving on the other sublattice,
which includes all the sites marked by circles (the analogs of O sites
in a MnO$_2$ lattice). The spin of the charge carrier is coupled through
exchange to the spins on its two nearest neighbor sites.

Such a model would describe for example a MnO$_2$ layer of the parent
compound LaMnO$_3$ if it was modeled as a charge-transfer
insulator.\cite{Fuj} The Mn ions are in the $3d^4$ ($S=2$)
configuration, and 
have ferromagnetic order in each layer.  In a charge-transfer
insulator, doping would 
introduce a hole in the O 
$2p$ orbitals instead of emptying an $e_g$ Mn $3d$ orbital. The spin
of the hole would be AFM coupled to the neighboring Mn spins, with a
$J_0 \sim |t_{p-d}|^2/\Delta E$, where $t_{p-d}$ would measure the
hybridization of the orbitals, and $\Delta E$ would be the overall
energetic cost to move the hole to a Mn site. 

This model is also  reminiscent of the Cu0$_2$ layers of cuprate
parent compounds, which are charge transfer materials with holes going
into O $2p$ orbitals. In this
case, $S=1/2$ for the $3d^9$ configuration of the Cu in the insulator,
however of course these spins order antiferromagnetically, not FM as
assumed to be the case here. 

In any event, our primary motivation here is to exemplify the general
solution for these types of problems, and we chose 
this model Hamiltonian  because it is complex enough to
demonstrate the full power of our solution, yet simple enough not to
make the  notations too cumbersome.  It also unveils some very
interesting physics. The solution can be directly generalized to a
much wider class of similar problems, as discussed below, and can
therefore be used to describe realistic systems.

We assume a total of $N$ unit cells, with periodic boundary
conditions. In all our results we let $N\rightarrow \infty$.  Each
unit cell is indexed by a pair of integers $(i,j)$ and contains three
sites: a spin site located at $\vec{R}_{ij} = ia \hat{x} + ja \hat{y}$
and 2 inequivalent electron sites, denoted $1$ (for the site on the
$x$-rung, location $\vec{R}_{ij,1} = \vec{R}_{ij} + {a\over 2}
\hat{x}$) and $2$ (site on the $y$-rung, location $\vec{R}_{ij,2} =
\vec{R}_{ij} + {a\over 2} \hat{y}$).

The hopping between the various electron sites is also indicated in
Fig. \ref{fig1}. For simplicity we limit ourselves to
nearest-neighbor hopping -- generalizations are straightforward.  If
the orbitals occupied by the charge carrier are $p$-orbitals (as drawn),
then one must take $t_1=-t_2=t$, whereas if they are $s$-type orbitals
(not shown), one should choose $t_1=t_2=t$. In any event, we will use
general $t_1$ and $t_2$ values for the derivation. The difference
between a hole vs. an electron charge carrier is the sign of $t$:
$t>0$ for electrons, and $t<0$ for holes. At first sight one would
think that the sign of $t$ is irrelevant, since the model is
particle-hole symmetric. In fact, while the energetics is the same,
the sign is important for the wavefunctions and therefore has
interesting consequences.

In the following, we assume that the charge carrier is an electron,
and we will comment on the differences for holes where appropriate.

We introduce the Fourier-transformed operators
\begin{equation}
\EqLabel{1}
c^\dag_{\vec{k},\lambda,\sigma}= {1\over \sqrt{N}} \sum_{i,j}^{}
e^{i\vec{k} \vec{R}_{ij,\lambda}}c^\dagger_{ij,\lambda,\sigma}
\end{equation}
where $c^\dagger_{ij,\lambda,\sigma}$ creates an electron with spin
$\sigma$ at site
$\lambda=1,2$ of the $(i,j)$ unit cell. In terms of these, the
nearest-neighbor hopping of the electron is written simply as:
\begin{equation}
\EqLabel{1a}
\hat{T} = \sum_{\vec{k},\sigma}^{}\epsilon_{\vec{k}}
\left(c^\dag_{\vec{k},1,\sigma} 
c_{\vec{k},2,\sigma} + h.c.\right) 
\end{equation}
where the sum is over the allowed values of $\vec{k}$ in the
first Brillouin zone (BZ) $(-{\pi\over a}, {\pi\over a}] \times (-{\pi\over
	a}, {\pi\over a}]$, and
\begin{equation}
\EqLabel{2}
\epsilon_{\vec{k}} = - 2t_1 \cos {(k_x-k_y)a\over 2} - 2 t_2 \cos
		{(k_x+k_y)a\over 2}.
\end{equation}
The hopping Hamiltonian is diagonalized trivially by using the new operators 
\begin{equation}
\EqLabel{3}
c_{\vec{k},\pm,\sigma} = {1\over \sqrt{2}} \left(c_{\vec{k},1,\sigma}
\pm c_{\vec{k},2,\sigma} \right),
\end{equation}
in terms of which the hopping Hamiltonian is:
\begin{equation}
\EqLabel{3a}
\hat{T} = \sum_{\vec{k},\sigma}^{} \epsilon_{\vec{k}}
\left(c^\dag_{\vec{k},+,\sigma} 
c_{\vec{k},+,\sigma} - c^\dag_{\vec{k},-,\sigma}
c_{\vec{k},-,\sigma}\right).
\end{equation}

As expected for this two-site unit cell, there are two bands for the
free electron (they happen to touch, there is no gap between them).
If $t_1=t_2$ (s-orbitals), the ground-state is at the $\Gamma$-point
$k_x=k_y=0$. If $t_1=-t_2$ (p-orbitals), the ground-state is moved to
$k_x=k_y=\pi/a$ or equivalent points at the corners of the Brillouin
zone. In fact, the whole dispersion is just translated by ${\pi\over
a} (1,1)$ inside the Brillouin zone, hence the resulting physics is
exactly the same in both cases. Differences between $s$- and
$p$-orbitals (besides this overall shift) are only apparent if there
is longer range hopping. Since we do not consider longer-range hopping
here, in the following we will report results for the $s$-orbital case
$t_1=t_2$.

To the  hopping term, we add the FM Heisenberg exchange
between the spins of magnitude $S$ :
\begin{equation}
\EqLabel{4}
{\cal H}_{\rm spins}=-J \sum_{i,j}^{} \left[\vec{S}_{i,j}
  \vec{S}_{i,j+1} + 
  \vec{S}_{i,j} \vec{S}_{i+1,j} - 2S^2 \right] 
\end{equation}
where the sum runs over all units cells 
(we limit ourselves to nearest-neighbor exchange since
generalizations to longer-range exchange are trivial, so long as there
is no
frustration) and the electron-spin exchange:
\begin{eqnarray}
\nonumber  {\cal H}_{\rm exc}= J_0 \sum_{i,j}^{} \left[\vec{s}_{ij,1}\cdot
 \left(\vec{S}_{ij} + \vec{S}_{i+1,j}\right)\right.& &\\
\left.+ \vec{s}_{ij,2}\cdot
 \left(\vec{S}_{ij}+ \vec{S}_{i,j+1}\right)\right] & &\EqLabel{5}
\end{eqnarray}
where $\vec{s}_{ij,\lambda}= \sum_{\alpha,\beta}^{}
c^\dagger_{ij,\lambda, \alpha} {\vec{\sigma}_{\alpha\beta}\over 2}
c^\dagger_{ij,\lambda, \beta}$ is the spin of the electron at site
$\lambda=1,2$ of the $(ij)$ unit cell. 

The total Hamiltonian is the sum of the three terms of
Eqs. (\ref{3a}), (\ref{4}) and (\ref{5}). For later convenience, we
divide it in two parts:
\begin{equation}
\EqLabel{5a}
{\cal H} = {\cal H}_0 + V 
\end{equation}
where 
\begin{equation}
\EqLabel{6b}
{\cal H}_0 = \hat{T} - J \sum_{i,j}^{} \left[{S}^z_{i,j}
  {S}^z_{i,j+1} + 
  {S}^z_{i,j}{S}^z_{i+1,j} - 2S^2 \right] 
\end{equation}
includes the hopping and the diagonal ($zz$) part of the FM
 Heisenberg interaction between the spins. $V$ includes the remaining
 terms, {\it i.e.} the $xy$ part of the FM exchange between spins, and the
coupling of the electron spin to the spins located on its two 
neighboring sites. 

\section{The Green's function of the spin-polaron}

We consider the case where a single electron is  in the
system and $T=0$. Let  $|{\rm FM}\rangle = |+S,+S,\dots,+S\rangle$ be
the ground-state of the spins, in the absence of the electron. If the
electron has spin-up the problem is trivially solved, since the $xy$
parts of all exchanges have zero action in this subspace (no spins can
be flipped). We therefore only treat explicitly the case of a
spin-down electron. 

We define
\begin{equation}
\EqLabel{4c}
\Psi_{\vec{k},\sigma} = 
\left(
\begin{array}[c]{c}
c_{\vec{k},1,\sigma} \\
c_{\vec{k},2,\sigma} \\
\end{array}
\right)
\end{equation}
and introduce the $2\times2$ Green's function matrix:
\begin{equation}
\EqLabel{8}
\bar{G}(\vec{k},\omega) = \langle {\rm FM}| \Psi_{\vec{k},\downarrow}
\hat{G}(\omega) \Psi^\dag_{\vec{k},\downarrow}|{\rm FM}\rangle,
\end{equation}
where the resolvent is $\hat{G}(\omega)=1/( \omega - {\cal
	H}+i\eta)$, with $\eta >0$  infinitesimally small, and we
	hereafter set
	$\hbar =1$. 
This shorthand notation means, for example, that the (1,2) element of
this matrix is:
$$
G_{1,2}(\vec{k},\omega) = \langle {\rm FM}| c_{\vec{k},1,\downarrow}
\hat{G}(\omega)c^\dag_{\vec{k},2,\downarrow} |{\rm FM}\rangle 
$$
etc., so that all four possible combinations are considered at
once. Using a Lehman representation,\cite{Mahan} it is clear that
$$
G_{\lambda,\lambda'}(\vec{k},\omega) = \sum_{u}^{} \frac{\langle {\rm
	FM}| c_{\vec{k},\lambda,\downarrow} |u\rangle\langle u|
  c^\dag_{\vec{k},\lambda',\downarrow} |{\rm FM}\rangle }{\omega - 
  E_u+ i\eta}
$$
where ${\cal H}|u\rangle = E_u |u\rangle$ are the single-electron
eigenstates in the sector of total $z$-axis spin $NS-{1 \over2}$. In
other words, the poles of these quantities give all the eigenenergies
and the residues measure overlaps between the true eigenfunctions and
the appropriate free-electron state
$c^\dag_{\vec{k},\lambda,\downarrow} |{\rm FM}\rangle, \lambda=1,2$.

Our goal is to calculate exactly this Green's function matrix. We do
this by using repeatedly Dyson's identity
$\hat{G}(\omega)=\hat{G}_0(\omega)+\hat{G}(\omega)V\hat{G}_0(\omega)$,
where $\hat{G}_0(\omega)=1/( \omega - {\cal H}_0+i\eta)$ and ${\cal
H}_0$ is defined in Eq. (\ref{6b}).

Rotating to the diagonal basis $c_{\vec{k},\pm,\downarrow}$ [see
  Eq. (\ref{3})] and back, it is straightforward to show that:
\begin{equation}
\EqLabel{9}
\hat{G}_0(\omega) \Psi^\dag_{\vec{k},\downarrow}|{\rm FM}\rangle =
\Psi^\dag_{\vec{k},\downarrow}|{\rm FM}\rangle  \bar{G}_0(\vec{k},\omega)
\end{equation}
where we introduce the $2\times2$ matrix:
\begin{equation}
\EqLabel{10}
\bar{G}_0(\vec{k},\omega) =
\left(
\begin{array}[c]{cc}
G_0^{(+)}(\vec{k},\omega) & G_0^{(-)}(\vec{k},\omega)\\
G_0^{(-)}(\vec{k},\omega) & G_0^{(+)}(\vec{k},\omega)\\
\end{array}
\right)
\end{equation}
in terms of the known free-electron propagators:
\begin{equation}
\EqLabel{11}
G_0^{(\pm)}(\vec{k},\omega)={1\over 2} \left[{1\over \omega -
	\epsilon_{\vec{k}} + i\eta} \pm {1\over \omega +
	\epsilon_{\vec{k}} + i\eta} \right].
\end{equation}

Using Dyson's identity once thus leads to the equation:
$$ \bar{G}(\vec{k},\omega) = \left[1 + \langle {\rm FM}|
\Psi_{\vec{k},\downarrow} \hat{G}(\omega) V
\Psi^\dag_{\vec{k},\downarrow}|{\rm
FM}\rangle\right]\bar{G}_0(\vec{k},\omega).
$$ Since the objects appearing here are $2\times2$ matrices, the order
of multiplications in these equations is important.

The action of $V$ on $\Psi^\dag_{\vec{k},\downarrow}|{\rm FM}\rangle$
  is easily estimated. The $xy$ exchange between spins has no
  contribution, since all spins are up. As a result, one finds
  contributions only due to the electron-spins exchange, resulting in:
\begin{eqnarray}
\nonumber \bar{G}(\vec{k},\omega) = \bar{G}_0(\vec{k},\omega)
\hspace{55mm}&& \\
\label{eq1}  + \left[-J_0S \bar{G}(\vec{k},\omega)+
   {J_0\over N} \sum_{\vec{q}}^{} \bar{F}(\vec{k},\vec{q},\omega)
  \bar{g}(\vec{q})\right] \bar{G}_0(\vec{k},\omega) \hspace{5mm} &&
\end{eqnarray} 
 The first term in the square bracket comes from the diagonal $zz$
exchange and corresponds to a simple shift in the total electron
energy (see below). The second term comes from the $xy$ exchange,
which allows one spin to be lowered by 1, while the electron-spin is
flipped to $\sigma=\uparrow$. This leads to a new $2\times 2$ Green's
function matrix:
$$
\bar{F}(\vec{k},\vec{q},\omega) = \sum_{i,j}^{}
e^{i\vec{q}\vec{R}_{ij}} \langle {\rm FM}| \Psi_{\vec{k},\downarrow}
\hat{G}(\omega) \Psi^\dag_{\vec{k}-\vec{q},\uparrow} S_{ij}^-|{\rm
FM}\rangle.$$ Finally, the diagonal matrix:
\begin{equation}
\EqLabel{13}
\bar{g}(\vec{q})=\left(
\begin{array}[c]{cc}
\cos{q_x a\over 2} & 0\\
0 & \cos {q_ya\over2}\\
\end{array}
\right)
\end{equation}
appears in Eq. (\ref{eq1}) because the spin at site $(ij)$ can be
flipped by exchange with an electron present at either of the
$(ij,1),(ij,2), (i-1,j,1)$ or $(i,j-1,2)$ sites, which are displaced
by $\pm a\hat{x}/2$ or $\pm a\hat{y}/2$ from it [see Eq. (\ref{1}) for
the definition of the phases in the Fourier transforms].

Given the structure of the $\bar{G_0}$ matrix, it follows that $
\left[\bar{G_0}(\vec{k}, \omega)\right]^{-1} + J_0S = 
\left[\bar{G_0}(\vec{k}, \omega+ J_0S)\right]^{-1}$, so we can further
simplify Eq. (\ref{eq1})  to:
\begin{equation}
\EqLabel{14}
\bar{G}(\vec{k},\omega) = \left[1 +
  {J_0\over N} \sum_{\vec{q}}^{} 
  \bar{F}(\vec{k},\vec{q},\omega)  \bar{g}(\vec{q})\right]
\bar{G}_0(\vec{k},\omega+J_0S).
\end{equation}
This shows that an equation for $\bar{F}$ is needed to solve the
problem. Using Dyson's identity again, we  find, after very similar kinds of
manipulations, that its equation of motion is:
\begin{eqnarray}
\nonumber \bar{F}(\vec{k},\vec{q},\omega) = 2J_0
 S \bar{G}(\vec{k},\omega)  \bar{g}(\vec{q})
 \bar{G}_0(\vec{k}-\vec{q},\omega-\Omega_{\vec{q}} -J_0S) &&\\ 
\EqLabel{16}-{J_0\over N} \sum_{\vec{Q}}^{}
\bar{F}(\vec{k},\vec{Q},\omega)
\bar{g}(\vec{Q}-\vec{q})\bar{G}_0(\vec{k}-\vec{q},\omega-\Omega_{\vec{q}}
-J_0S)\hspace{3mm}  &&
\end{eqnarray}
where
\begin{equation}
\EqLabel{17}
\Omega_{\vec{q}} = 4JS\left(\sin^2
	  {q_xa\over 2} +  \sin^2
	  {q_ya\over 2} \right) 
\end{equation}
is the one-magnon spectrum for this FM spin lattice.

Eqs. (\ref{14}) and (\ref{16}) can now be solved to find
$\bar{G}(\vec{k},\omega)$ and $\bar{F}(\vec{k},\omega)$. Note that
usually there is an infinite sequence of equations-of-motion connected
to one another. In this problem, the series is truncated to just 2
equations because of symmetries: since $\hat{S}_{z,{\rm tot}}$ (which
includes all spins and the electron) commutes with the Hamiltonian,
the evolution is always within the Hilbert sector with $z$-axis spin
$NS-1/2$. This only includes the states with all spins up and the
electron with spin-down; or the electron has spin-up, and then one
spin is lowered by one (one magnon is created in the system). This
also explains why generalizations to finite $T$ (multiple magnons) or
finite electron concentrations are far from trivial: in those cases,
more and more equations of motion are generated as the size of the
relevant Hilbert subspace increases substantially, and their solution
becomes very difficult.

Eqs. (\ref{14}) and (\ref{16}) can be solved analytically because of
the simple structure of the $\bar{g}$-matrix, which contains only
trigonometric functions.  In fact, if we also
define:
\begin{equation}
\EqLabel{17b}
\tilde{g}(\vec{q})=\left(
\begin{array}[c]{cc}
\sin{q_x a\over 2} & 0\\
0 & \sin {q_ya\over2}\\
\end{array}
\right)
\end{equation}
then we can ``factorize'':
\begin{equation}
\EqLabel{18}
\bar{g}(\vec{Q}-\vec{q}) = \bar{g}(\vec{Q})\bar{g}(\vec{q})+
\tilde{g}(\vec{Q})\tilde{g}(\vec{q}) 
\end{equation}
since 
$
\cos{(Q-q)a\over 2} = \cos {Qa\over 2} \cos {q a\over 2} +
\sin {Qa\over 2} \sin {q a\over 2}
$.

 We define the auxiliary quantities:
\begin{equation}
\EqLabel{19}
\bar{f} (\vec{k},\omega)= {1\over N} \sum_{\vec{Q}}^{}
\bar{F}(\vec{k},\vec{Q},\omega) \bar{g}(\vec{Q}),
\end{equation}
which is the only quantity we need to compute $\bar{G}$, see Eq. (\ref{14}),
and
\begin{equation}
\EqLabel{20}
\tilde{f} (\vec{k},\omega)= {1\over N} \sum_{\vec{Q}}^{}
\bar{F}(\vec{k},\vec{Q},\omega) \tilde{g}(\vec{Q}) 
\end{equation}
in terms of which we can rewrite Eq. (\ref{16}) as:
\begin{eqnarray}
\nonumber  &&\bar{F}(\vec{k},\vec{q},\omega) = J_0\left[-\bar{f}
  (\vec{k},\omega) \bar{g}(\vec{q}) -\tilde{f}
  (\vec{k},\omega) \tilde{g}(\vec{q})  \right.\\
\EqLabel{21} &&\left. +2S
  \bar{G}(\vec{k},\omega)  \bar{g}(\vec{q})    \right] 
\bar{G}_0(\vec{k}-\vec{q},\omega-\Omega_{\vec{q}} -J_0S)
\end{eqnarray}
Substituting this in Eqs. (\ref{19}) and (\ref{20}), we  obtain two
linear 
(matrix) equations with unknowns $\bar{f}$ and $\tilde{f}$, and
inhomogeneous terms proportional to $\bar{G}$:
\begin{widetext}
$$
\bar{f} (\vec{k},\omega)= \left[-J_0 \bar{f} (\vec{k},\omega) + 2
  J_0S\bar{G}(\vec{k},\omega)  
  \right] \bar{g}_{11}(\vec{k},\omega) -  J_0 \tilde{f}
(\vec{k},\omega)  \bar{g}_{21}(\vec{k},\omega) 
$$
and
$$
\tilde{f} (\vec{k},\omega)= \left[-J_0 \bar{f} (\vec{k},\omega) + 2
  J_0S\bar{G}(\vec{k},\omega)  
  \right] \bar{g}_{12}(\vec{k},\omega) - J_0 \tilde{f}
(\vec{k},\omega)  \bar{g}_{22}(\vec{k},\omega) 
$$
\end{widetext}
where we introduced the known 2$\times2$ matrices:
\begin{equation}
\EqLabel{22}
\bar{g}_{11}(\vec{k},\omega)= {1\over N} \sum_{\vec{q}}^{}
\bar{g}(\vec{q}) 
\bar{G}_0(\vec{k}-\vec{q},\omega-\Omega_{\vec{q}} -J_0S) 
\bar{g}(\vec{q}) 
\end{equation}
\begin{equation}
\EqLabel{23}
\bar{g}_{12}(\vec{k},\omega)= {1\over N} \sum_{\vec{q}}^{}
\bar{g}(\vec{q}) 
\bar{G}_0(\vec{k}-\vec{q},\omega-\Omega_{\vec{q}} -J_0S) 
\tilde{g}(\vec{q}) 
\end{equation}
\begin{equation}
\EqLabel{24}
\bar{g}_{21}(\vec{k},\omega)= {1\over N} \sum_{\vec{q}}^{}
\tilde{g}(\vec{q}) 
\bar{G}_0(\vec{k}-\vec{q},\omega-\Omega_{\vec{q}} -J_0S) 
\bar{g}(\vec{q}) 
\end{equation}
\begin{equation}
\EqLabel{25}
\bar{g}_{21}(\vec{k},\omega)= {1\over N} \sum_{\vec{q}}^{}
\tilde{g}(\vec{q}) 
\bar{G}_0(\vec{k}-\vec{q},\omega-\Omega_{\vec{q}} -J_0S) 
\tilde{g}(\vec{q}) 
\end{equation}
In fact, lots of these matrices' elements are related to each other by
various symmetries, 
so  fewer than $16$ actually need to be
calculated. Letting $N\rightarrow \infty$, then each of these
corresponds to a
two-dimensional integral over the Brillouin zone. One integral can be
performed analytically, and the second  we integrated numerically,
therefore these matrices are easy to calculate. 

These coupled equations are easy to solve  and the resulting expression of
$\bar{f}$ can now be used in Eq. (\ref{14}) to find the Green's
function explicitly. The final result is: 
\begin{widetext}
\begin{equation}
\EqLabel{27}
\bar{G}(\vec{k},\omega)  =
\left[\left(\bar{G}_0(\vec{k},\omega+J_0S)\right)^{-1} - 2S J_0
  \left(1- \left(\bar{M}(\vec{k},\omega)\right)^{-1} \right)\right]^{-1}
\end{equation}
where we introduced the matrix:
\begin{equation}
\EqLabel{26}
\bar{M}(\vec{k},\omega) = 1 + J_0 \left[\bar{g}_{11}(\vec{k},\omega) -
  J_0 \bar{g}_{12}(\vec{k},\omega)\left(1+ J_0
  \bar{g}_{22}(\vec{k},\omega)\right)^{-1} 
  \bar{g}_{21}(\vec{k},\omega)\right].
\end{equation}
\end{widetext}

Several comments are in order. First, the electron-spin exchange $J_0$
appears in three places in Eq. (\ref{27}), namely (i) in the overall
shift by $-J_0S$ of the energy argument of the $\bar{G}_0^{-1}$ term
(first term in the denominator); (ii) as an overall factor for the
``self-energy'' (second term in the denominator) and (iii) as a shift
by $+J_0S$ in the energy argument of the $\bar{G}_0$ functions
appearing in the definitions of $\bar{g}_{11}$, etc. in the
self-energy. The first and third of these are due to the $zz$
exchange, which simply shifts the energy of the electron by $\pm J_0S$
depending on whether its spin is parallel or antiparallel to the FM
background. The second is due to the $xy$ exchange, which facilitates
the spin-flip of the electron. Therefore, the generalization to an
anisotropic interaction is straightforward, for instance if
$J_{0,\perp}=\lambda J_{0,z}$ then the self-energy is multiplied by
$\lambda$. The exchange $J$ between spins appears only in the magnon
dispersion $\Omega_{\vec{q}}$, which enters only the $\bar{g}_{11}$,
etc. functions appearing in the self-energy. This is not surprising,
since that self-energy term is due to contributions from one-magnon
plus spin-up electron states. If the FM exchange between spins is
longer range, one simply has to replace the magnon dispersion by the
appropriate one.

Eq. (\ref{27}) thus reveals that the free spin-down electron state of
bare energy $\epsilon_{\vec{k}} -J_0S$ is mixed, through spin-flipping
due to the $xy$ term, with the continuum of one-magnon plus spin-up
electron states of bare energies $\epsilon_{\vec{k}-\vec{q}} +
J_0S+\Omega_{\vec{q}}$.  It follows that if we are interested in
having an infinitely long-lived quasiparticle state at low energies,
then the electron-spin exchange must be antiferromagnetic $J_0>0$
(this is easily confirmed numerically). We will focus on this
situation in the following (the $J_0<0$ case is less interesting as
the low-energy dynamics is incoherent, with finite lifetime
excitations).

Let us now briefly comment on other generalizations. The reason we
have equations for $2\times2$ matrices is that, in this problem, the
electron unit cell has a two-site basis. For a basis with $n$
different sites per unit cell one would have similar equations but for
$n\times n$ matrices, where all the electron hopping information would
be encoded in the corresponding $\bar{G}_0$ matrix. The dimensionality
of the problem enters only in the sums over the Brillouin zone, i.e. in
the number of integrals to be performed. Slightly more complicated is
the generalization of the electron-spin exchange to longer range. This
leads to the appearance of more auxiliary functions like $\bar{f}$ and
$\tilde{f}$, i.e. one needs to solve a linear system with more
unknowns in order to find the self-energy. The various auxiliary
functions correspond to the different possible phase-shifts (analogs
of the $\cos(q_{x,y}a/2), \sin(q_{x,y}a/2)$ appearing in the $\bar{g},
\tilde{g}$ matrices). For example, assume that for the same system
discussed here, the electron can visit all sites, including the spin
sublattice. Also, assume that there is on-site exchange between the
electron and the local spin, if the electron is on the spin
sublattice, besides the exchange discussed here. In that
case, one has to deal with $3\times 3$ matrices (three-site basis) and
there are three auxiliary functions, two similar to the ones that
appeared here, and one corresponding to zero phase-shift for the
on-site interaction (because of the zero phase-shift, there is no
auxiliary function proportional to $\sin(0)$). 

We have
checked explicitly that one can also include
more complicated terms, for 
example electron hopping accompanied by a spin-flip coupled to a spin
lowering of a nearby lattice spin, like
$c^\dagger_{ij,1,\uparrow}c_{ij,2,\downarrow} S^-_{ij}$ etc.
(such terms have important consequences, as we
discuss below). In fact, we believe that essentially any problem
from this class is solvable analytically, along these general lines.

\section{Results}

As already mentioned, we present results for the more interesting
case of an AFM electron-spin exchange, $J_0>0$, whose low-energy state
is an infinitely-lived spin-polaron. Also, the results  are
for $t_1=t_2=t$ ($s$-orbitals), and $t$ will be used as the energy
unit. As discussed previously, for the simple nearest-neighbor hopping
used here, the only difference for $p$-orbitals would be to shift the
values of all momenta by ${\pi\over a}(1,1)$. For holes, $t$ changes
sign with consequences discussed where appropriate.

\begin{figure}[b]
\includegraphics[width=0.9\columnwidth]{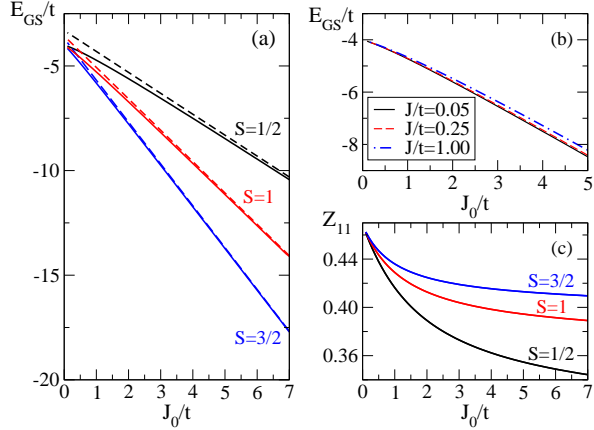}
\caption{(color online) (a) Spin-polaron ground-state energy, in units
  of $t$, as a function of $J_0/t$, for $S={1\over 2}, 1$ and ${3\over
  2}$,and $J/t=0.05$. The full lines show the exact results, while the
  dashed lines give the strong-coupling perturbational limit discussed
  in the text; (b) Spin-polaron ground-state energy, in units of $t$, as a
  function of $J_0/t$, for $S={1\over 2}$ and $J/t=0.05, 0.25$ and
  1.00; (c) Spectral weight $Z_{11}= |\langle {\rm FM}| c_{\vec{k}, 1,
  \downarrow}| {\rm GS}\rangle|^2$ as a function of $J_0/t$ for three values
  of $S$ and $J/t=0.05$. }
\label{fig2}
\end{figure}

We begin by analyzing the dependence of the ground-state energy
$E_{\rm GS}$ and quasiparticle weight on various parameters of the
problems. In panel (a) of Fig. \ref{fig2}, the full lines illustrate
the dependence of $E_{\rm GS}$ on the electron-spin exchange $J_0/t$,
as well as the spin value $S$, for a fixed spin-spin exchange
$J/t=0.05$. As expected, the GS energy is lowered as both $J_0$ and
$S$ increase, mainly due to the favorable $zz$-exchange between the
lattice of FM spins and the spin-down electron.

The dashed lines in Fig.~\ref{fig2}(a) show the asymptotic expressions
obtained by first order perturbation in the hopping $t$, in the
strong-coupling limit $J_0/t \rightarrow \infty$. The agreement is
very reasonable even for rather small $J_0/t$ values, therefore it is
useful to analyze this solution in some detail, to
understand the nature of the spin-polaron. 

In the absence of hopping,
$t=0$, we can form two sets of translationally-invariant states of
well-defined momentum $\vec{k}$, which are ground-states of the
electron-spin exchange ${\cal H}_{\rm exc}$, namely:
\begin{widetext}
\begin{eqnarray}
\EqLabel{b1}  &&|\vec{k},1\rangle = \sum_{i,j}^{}
{e^{i\vec{k}\cdot\vec{R}_{ij,1}}\over \sqrt{N}}\sqrt{1\over
  4S+1}\left[\sqrt{4S}c^\dagger_{ij,1,\downarrow} -{1\over \sqrt{4S}}
  c^\dagger_{ij,1,\uparrow}\left(S_{ij}^-+S_{i+1,j}^-\right)
  \right]|{\rm FM}\rangle \\
\EqLabel{b1b}  &&|\vec{k},2\rangle = \sum_{i,j}^{}
{e^{i\vec{k}\cdot\vec{R}_{ij,2}}\over \sqrt{N}}\sqrt{1\over
  4S+1}\left[\sqrt{4S}c^\dagger_{ij,2,\downarrow} -{1\over \sqrt{4S}}
  c^\dagger_{ij,2,\uparrow}\left(S_{ij}^-+S_{i,j+1}^-\right)
  \right]|{\rm FM}\rangle
\end{eqnarray}
\end{widetext}
For all of these states ($\lambda=1,2$), 
$${\cal H}_{\rm exc} |\vec{k},\lambda\rangle =
-J_0\left(S+{1\over 2}\right) |\vec{k},\lambda\rangle. $$

The hopping $\hat{T}$ lifts this degeneracy. In fact, it is
straightforward to show that $\langle \vec{k}, \lambda|
\hat{T}|\vec{k}',\lambda\rangle =0$ while $\langle \vec{k}, 1|
\hat{T}|\vec{k}',2\rangle = \delta_{\vec{k},\vec{k}'} {4S+{1\over 2}\over
  4S+1}\epsilon_{\vec{k}}$. Thus, within this level of perturbation
theory, the eigenstates are 
\begin{equation}
\EqLabel{b4}
|\vec{k},\pm\rangle = {1\over \sqrt{2}} \left(|\vec{k},1\rangle  \pm
|\vec{k},2\rangle \right) 
\end{equation}
and their corresponding energy is found to be:
\begin{equation}
\EqLabel{b5}
E_{\vec{k},\pm}^{(p)} = \pm \epsilon_{\vec{k}} \cdot {4S+{1\over 2}\over
  4S+1} + {3JS\over 4S+1} - J_0\left(S+{1\over 2}\right).
\end{equation}
The middle term is due to the spin-spin FM exchange energy lost
because part of the wavefunction has one spin in the background
lowered by 1.

It follows that the perturbational prediction for the ground-state
energy, shown in Fig. \ref{fig2}, is:
\begin{equation}
\EqLabel{b6} E_{GS}^{(p)} = -4t {4S+{1\over 2}\over 4S+1}+ {3JS\over
4S+1} - J_0\left(S+{1\over 2}\right)
\end{equation}
In the limit of large $J_0/t$ the agreement with the exact solution is
very good. In the limit $J_0\rightarrow 0$, and ignoring the small
correction proportional to $J$ (which is reasonable, given the small
$J/t$ value used), we see that $E_{GS}^{(p)} \rightarrow -4t
{4S+{1\over 2}\over 4S+1}> -4t$, whereas the true GS energy cannot go
above $-4t$ (this is the free-electron GS energy). For large $S$ the
difference between these two values becomes negligible, however for
$S={1\over 2}$ the difference is sizable, explaining the significant
difference between the two curves in the low-$J_0$ part of Fig. \ref{fig2}(a).

In Fig. \ref{fig2}(b) we show the exact GS energies for different
values of $J/t$ and $S={1\over 2}$. In the limit $J_0/t \rightarrow 0$
there is no interaction between the electron and the FM background,
therefore the value of $J$ is irrelevant. As expected from the
discussion above, at large $J_0/t$ values the GS energy increases
linearly with increasing $J$. However, this is a small contribution
for reasonably small values of $J$. In the following we fix $J=0.05t$.

In Fig. \ref{fig2}(c) we plot the quasiparticle ({\em qp}) weights in
the ground-state, defined as:
\begin{equation}
\EqLabel{b7} Z_{\lambda\lambda}= |\langle {\rm FM}| c_{\vec{k}=0, \lambda,
\downarrow}| {\rm GS}\rangle|^2.
\end{equation}
Clearly, these quantities give the probability to find the electron
with spin-down on the sublattice $\lambda=1,2$, in the GS. As expected
by symmetry, $Z_{11}=Z_{22}$ therefore only one is shown in
Fig. \ref{fig2}(c), as a function of $J_0/t$. One immediate
observation is that the {\em qp} weights saturate to finite values in
the strong coupling limit $J_0/ t\rightarrow \infty$, instead of
becoming exponentially small, as is the case for typical polarons
(where the electron binds phonons in its vicinity, due to
electron-phonon coupling).\cite{MB1} The reason, of course, is that
here the electron can bind a maximum of one magnon as opposed to an
arbitrarily large number of phonons, as is the case with polarons for
increasing electron-phonon coupling.

By direct comparison of Eqs. (\ref{b1})-(\ref{b4}) and (\ref{1}),
(\ref{3}), it is clear that in the strong coupling limit, the
spin-down part of the low energy $|\vec{k},+\rangle $ spin-polaron
state is equal to $\sqrt{4S/(4S+1)}
c^\dagger_{k,+,\downarrow}|{\rm FM}\rangle $. In other words, for any
momentum $\vec{k}$ of the spin-polaron, the probability to find the
electron with spin-down and the FM background undisturbed is
$4S/(4S+1)$. By symmetry, it follows that $Z_{11}=Z_{22}= 2S/(4S+1)$
for all spin-polaron momenta, including the GS. This is in reasonable
agreement with the values shown in Fig. \ref{fig2}(c), given that even
$J_0/t=7$ is not that large, therefore corrections beyond first order
perturbation are not expected to be negligible.

The spin-flipped contribution to the $|\vec{k},+\rangle $ low-energy
spin-polaron 
state also reveals very interesting physics. Focusing on the case
$S=1/2$, we can rewrite the spin-polaron eigenstates in the strong
coupling limit as:
\begin{equation}
\EqLabel{b8} |\vec{k}, +\rangle = {1\over \sqrt{12 N}} \sum_{ij}^{}
e^{i\vec{k}\cdot \vec{R}_{ij}}
\left[\chi^\dagger_{ij,\downarrow}(\vec{k}) -S^-_{ij}
  \chi^\dagger_{ij,\uparrow}(\vec{k})\right]|{\rm FM}\rangle.
\end{equation}
The operator in the bracket is recognized as creating a singlet
between the spin located at site $ij$ and the electron occupying a
state centered at site $ij$. This singlet propagates with momentum
$\vec{k}$ through the  FM background. The electron state that
forms the singlet with the spin is found to be
$\chi^\dagger_{ij,\sigma}(\vec{k}) = e^{i{k_x a\over 2}}
c^\dagger_{ij,1, \sigma} + e^{i{k_y a\over 2}} c^\dagger_{ij,2,
\sigma} + e^{-i{k_x a\over 2}} c^\dagger_{i-1,j,1, \sigma} + e^{-i{k_y
a\over 2}} c^\dagger_{i,j-1,2, \sigma}$, {\em i.e.} a superposition of
the four electronic sites surrounding the spin located at $ij$ (site
labeling is shown in Fig. \ref{fig1}). For larger $S$ the solution is
analogous, except one cannot speak of a ``singlet'' between a
spin-${1\over 2}$ and a spin $S>{1\over 2}$. However, it can be shown
that the entangled electron-spin state corresponds to a total spin
$S-1/2$ (not surprising, since this is total spin that minimizes the
AFM exchange energy).

In the ground-state, the coefficients are determined by the orbitals
participating in the hopping. For our model we  find for $\vec{k}=0$ that
$\chi^\dagger_{ij,\sigma}= c^\dagger_{ij,1, \sigma} + c^\dagger_{ij,2,
\sigma} + c^\dagger_{i-1,j,1, \sigma} + c^\dagger_{i,j-1,2, \sigma}$,
while for $p$-orbitals a similar calculation (or see
$\vec{k}={\pi\over a} (1,1)$ case) leads to $\chi^\dagger_{ij,\sigma}=
-c^\dagger_{ij,1, \sigma} -c^\dagger_{ij,2, \sigma} +
c^\dagger_{i-1,j,1, \sigma} + c^\dagger_{i,j-1,2, \sigma}$. In other
words, the signs mirror the sign of the lob
pointing towards the central spin site. 

If the charge carrier was a hole instead of an electron, because
$t\rightarrow -t$
the low-energy spin-polaron eigenstate is  $|\vec{k}, -\rangle$
(since now $\epsilon_{\vec{k}} >0$). The eigenenergy is the same, but
the orbital  
involved in the singlet is  $\chi^\dagger_{ij,\sigma}(\vec{k}) =
e^{i{k_x a\over 2}} 
c^\dagger_{ij,1, \sigma} - e^{i{k_y a\over 2}} c^\dagger_{ij,2,
\sigma} + e^{-i{k_x a\over 2}} c^\dagger_{i-1,j,1, \sigma} - e^{-i{k_y
a\over 2}} c^\dagger_{i,j-1,2, \sigma}$. As a result, for $s$-orbitals
the GS orbital becomes
$\chi^\dagger_{ij,\sigma}= c^\dagger_{ij,1, \sigma} - c^\dagger_{ij,2, 
\sigma} + c^\dagger_{i-1,j,1, \sigma} - c^\dagger_{i,j-1,2,
  \sigma}$, {\em i.e.} it has $d$-like symmetry. For $p$-orbitals it
 has $p$-like symmetry again, but is orthogonal to the one listed above
for the electron.

In conclusion, the particular linear combination selected for forming
the GS singlet (more generally, $S-{1\over 2}$ state) with the central
spin is determined both by the particular orbitals involved, and
 by the nature -- hole or electron -- of the charge carrier.

This solution is clearly analogous to the Zhang-Rice (ZR)
singlet~\cite{ZR} but with some differences. For one, the singlet
defining the spin-polaron is here propagating in a FM, not AFM
background. Second, while the ZR singlet involves a $d$-wave like linear
combination of electronic orbitals, here  the combination depends on
the details of the model considered, as just discussed. 

The bigger difference, of course, is that here we have spins at the
Mn-like sites. Of course, spins arise from having some atomic orbital
partially filled, and one can talk about a well defined spin when the
number of electrons (holes) in this shell cannot change. As noted above, we
can easily generalize our model to allow the extra electron (hole) to hop
onto the Mn-like sites, adding a Hubbard-$U$ penalty and/or Hundt's
exchange as well, if
desired. What we cannot do, at least so far, is solve exactly the more
general model where electrons (holes) that are currently locked onto the
Mn-like sites and constitute their spins, are allowed to hop to the
other sublattice, so that a spin less than $S$ is left behind. The
difficulty is simple to see: even if one adds a charge-gap $\Delta$,
{\em i.e.} an energy penalty to move electrons (holes) from the
Mn-like sites to the O-like sites, in any eigenstate there will be
some finite probability to find any number of $O$-like sites occupied
and the wavefunctions become too complicated. To be more precise, one
can still find easily the equivalent of the FM-background state in
this case, i.e. the ground-state in the Hilbert sector of z-axis spin
$NS$. The case of spin $NS+n{1\over2}$, when any number $n$ of
electrons (holes) with spin parallel to the background have been added in, is
also trivially solved (because Hubbard on-site repulsion does not act
between parallel spins, therefore both cases are essentially without
interactions). However the problem corresponding to $NS-1/2$, i.e. for
adding a spin-down electron (hole) to the FM ``background'', in other words
the equivalent of the simpler problem investigated here, has proved
too complicated for us so far.

However, even the asymptotic limit of our simplified model still
provides a very important insight, namely that the phases of the
electronic orbitals locked in the singlet with the central spin {\em
vary inside the Brillouin zone}. In the ZR model\cite{ZR} these phases
are assumed to be locked to their GS, $d$-wave symmetry values
irrespective of the momentum of the ZR singlet. As is well known, that
leads to problems with normalization of the resulting states in some
regions of the BZ. What our simpler but exact solution reveals is that this is not
correct: the phases of the orbitals involved in the singlet vary at
different $\vec{k}$-points. This insight might help improve the
description of the ZR singlet away from the $\vec{k}=0$ region.

In terms of $\vec{k}$-dependent properties, of course we can extract
not only the GS state, but the entire dispersion of the spin-polaron,
by focusing on the lowest (discrete) eigenstate of momentum
$\vec{k}$. In standard polaron physics, the polaron bandwidth is
expected to become smaller (corresponding to larger effective polaron
mass) as the electron-phonon coupling increases and more phonons are
tied into the polaron cloud.\cite{MB1} Here we do not expect to have
this problem since a maximum of one magnon can be bound to the
electron, as already discussed. In fact, Eq. (\ref{b5}) reveals that
the spin-polaron mass becomes, in the strong-coupling asymptotic
limit:
\begin{equation}
\EqLabel{b10} {m^*\over m}= {4S+1\over 4S+{1\over 2}}
\end{equation}
where $m$ is the bare electron mass. This is interesting as it is
determined only by the value $S$ of the spins in the FM
background, and independent of the coupling. The largest increase of
$6/5=1.2$ is for spins $S=1/2$, 
showing that the spin-polaron remains very light. This is not
surprising, given its propagation in an FM background.

\begin{figure}[t]
\includegraphics[width=0.9\columnwidth]{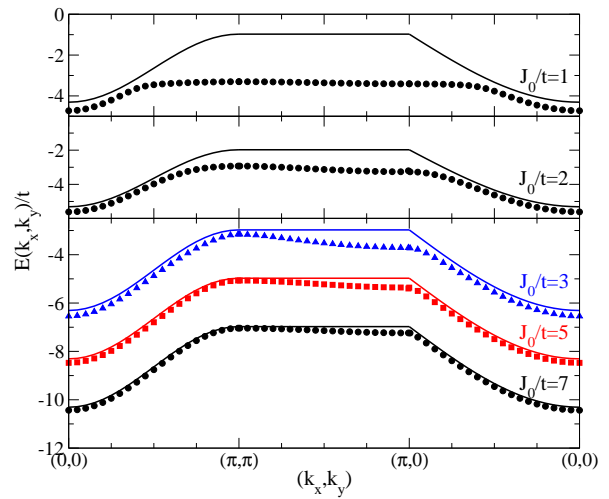}
\caption{(color online) Spin-polaron dispersion along lines of high
  symmetry in the Brillouin zone. The results (symbols) correspond to
  $S={1\over 2}$ and $J/t=0.05$ and, from top to bottom,
  $J_0/t=1,2,3,5$ and 7. The top two cases have been plotted separately to
  avoid overlaps. The lines show the strong-coupling prediction
  $E^{(p)}_{\vec{k},+}$ of Eq. (\ref{b5}).}
\label{fig3}
\end{figure}

In Fig. \ref{fig3} we plot the spin-polaron dispersion (lowest
eigenstate of momentum $\vec{k}$) along  lines of high symmetry
inside the BZ. The full lines show the strong-coupling prediction
$E^{(p)}_{\vec{k},+}$ of Eq. (\ref{b5}) -- again, we assume
$s$-orbitals and an electron as the charge carrier. The agreement is very good
for larger $J_0/t$ values, suggesting that the paradigm of the spin
polaron as a mobile singlet of the electron in a special state and the
spin at its center, must be a good description even for moderate
values of $J_0/t$. Even for $J_0/t=1$ the agreement is quite
reasonable near the center of the BZ, but not at high $\vec{k}$
values. This discrepancy can be understood easily as well. As
discussed, the spin-polaron states come from hybridizing the
non-interacting spin-down electron states of energy
$\epsilon_{\vec{k}}-J_0S$ with the spin-up electron + one magnon
continuum of energies $\epsilon_{\vec{k}-\vec{q}}
+J_0S+\Omega_{\vec{q}}$. The hybridization will push the discrete
state to lower energies, giving rise to the spin-polaron band, however
it cannot change the location of the continuum. If we ignore the
$\Omega_{\vec{q}}$ term, given the small $J/t$, then we find that this
continuum must start at $-4t+J_0S$ (inclusion of $\Omega_{\vec{q}}$
adds a small, $\vec{k}$-dependent correction to this value of the
continuum band-edge, see results below). The spin-polaron discrete
state cannot overlap 
with the continuum, therefore when $J_0$ is small, one expects the
polaron dispersion to flatten out just below the continuum band-edge,
which is precisely what we see for $J_0/t=1$. As $J_0/t$ increases the
continuum moves to higher energies while the spin-polaron band moves to
lower energies and becomes fully visible.

The flattening-out of the polaron dispersion just under the continuum
is typical polaron physics at weak coupling.\cite{MB1} Of course, for
conventional polarons the continuum starts at $\Omega$ above the
polaron ground-state, where $\Omega$ is the energy of the Einstein
phonons. This limits the polaron bandwidth, at weak coupling, to be
precisely $\Omega$. For the spin-polaron studied here, the bandwidth
has little to do with the magnon energy (the FM magnons are, in fact,
gapless).  Instead, the relevant energy scale comes from the $zz$
component of the exchange energy between the electron and the FM
background, $J_0S$.

\begin{figure}[t]
\includegraphics[width=0.7\columnwidth]{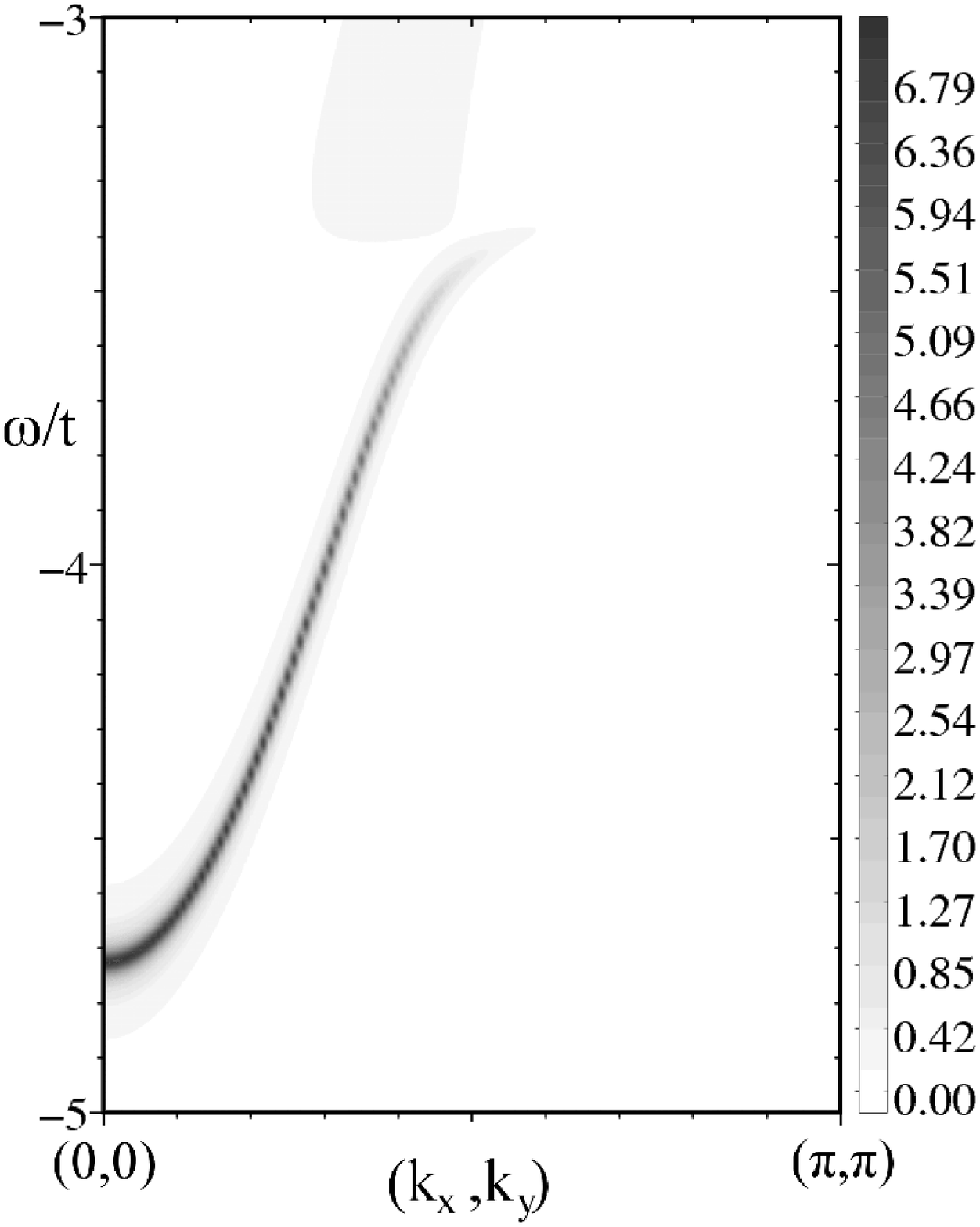}
\includegraphics[width=0.7\columnwidth]{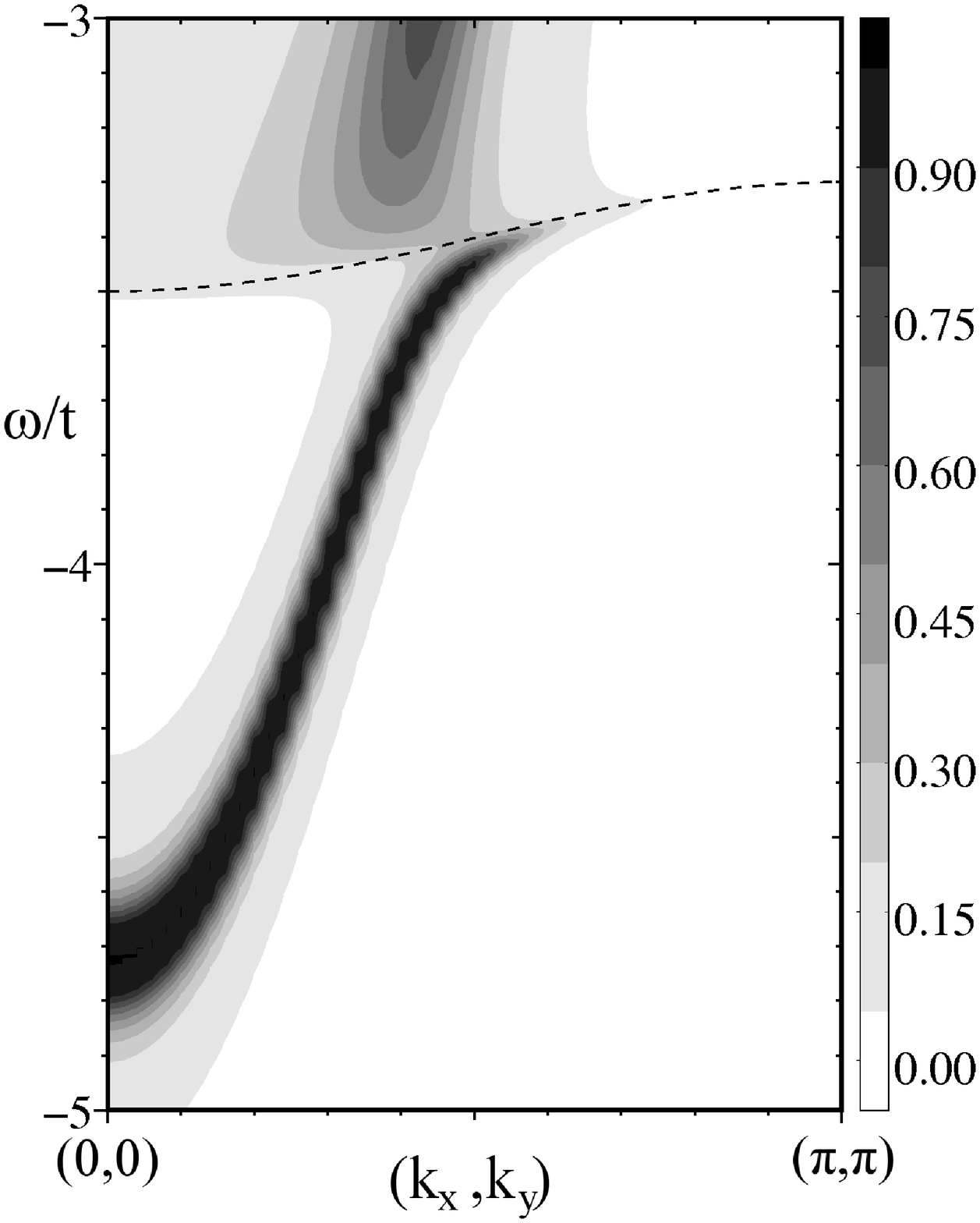}
\caption{Top panel: Contour plot of $A_{11}(kx,ky,\omega)$ along
the  $k_x=k_y$ cut in the Brillouin zone. Only the spin-polaron state
is visible on this scale. Bottom panel: same as above, but plotted as
$\tanh[A_{11}(kx,ky,\omega)/0.3]$ so as to make low-weight features
visible. The dashed line shows the 
  expected location of the lower edge of the continuum. Parameters are
  $J_0=t=1, J=0.05, S=0.5$ and $\eta = 0.02$.}
\label{fig4}
\end{figure}

Another typical expectation for  polaron physics at weak couplings
is that of very small spectral weight in the BZ regions where the
polaron dispersion lies just below the continuum. This is because here the
largest contribution to the polaron comes from continuum states where
the electron is spin-up. Consequently, the probability to find a
spin-down electron, measured by the $qp$ weights, becomes very
small. This is indeed confirmed in Figs. \ref{fig4}, where we show the
spectral weight
\begin{equation}
\EqLabel{a11} A_{\lambda\lambda}(\vec{k},\omega) = -{1\over \pi} {\rm
Im} G_{\lambda\lambda}(\vec{k},\omega)
\end{equation}
along the $k_x=k_y$ cut in the BZ. Again, by symmetry we expect
$A_{11}=A_{22}$. The top panel shows the spectral weight on a linear
scale. The low-$k$ part of the spin-polaron band is clearly visible,
however the region where it flattens just under the continuum has very
little weight, and is not visible on this scale. 
In order to make low-weight features near the
continuum more visible, instead of $A_{11}(\vec{k},\omega)$, in the
bottom panel we plot
$\tanh[ A_{11}(\vec{k},\omega)/0.3]$. As a result, all regions with
weight $A_{11} > 0.7$ are mapped into black, whereas weights below
this follow a fairly linear scale down to white. This explains why the
spin-polaron peak seems so wide now, although in reality it is a
Lorentzian of width $\eta\rightarrow 0$. The dashed line shows the
continuum band-edge, $\min_{\vec{q}} [ \epsilon_{\vec{k}-\vec{q}} +
J_0S+\Omega_{\vec{q}}]$. As expected, the spin-polaron dispersion
flattens out just underneath it. Because its weight decreases so fast
as $\vec{k}$ increases, it is impossible to see it even on this scale
for larger $\vec{k}$. The continuum above is also more clearly seen for
smaller values of $\vec{k}$, with most weight around
$\epsilon_{\vec{k}} - J_0S$ where the free spin-down electron would
have its bare energy. The little weight that seems to ``seep'' below the
dashed line at low-$\vec{k}$ is due to the finite $\eta$ used, we have
checked that indeed the continuum appears at the expected value.

\begin{figure}[t]
\includegraphics[width=0.8\columnwidth]{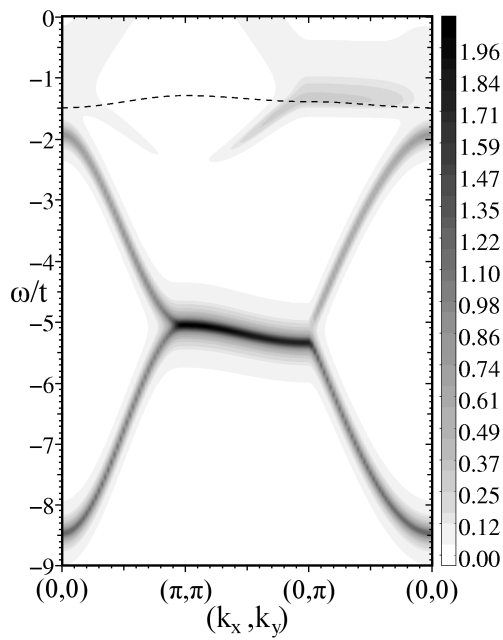}
\caption{Contour plot of $A_{11}(k_x,k_y,\omega)$ along
  several cuts in the Brillouin zone. The dashed line shows the
  expected location of the lower edge of the continuum. Parameters are
  $t=1, J_0=5, J=0.05, S=0.5$ and $\eta = 0.1$.}
\label{fig5}
\end{figure}

As $J_0$ increases and the spin-polaron band moves well below the
continuum, we expect to see large $qp$ weight for the spin-polaron at
all $\vec{k}$ since the strong-coupling limit predicts a
$\vec{k}$-independent $qp$ weight. This is indeed the case, as shown
in Fig. \ref{fig5} for $J_0/t=5$. Note that the spectral weight is now
shown on a linear scale. The low-energy polaron band has the same
dispersion as shown in Fig. \ref{fig3}, and indeed it has a fairly
constant $qp$ weight. One initial surprise is to see another coherent,
infinite lifetime state just above it (and in fact degenerate with it
along the $(\pi,\pi)-(\pi,0)$ cut, leading to a doubled $qp$ weight there),
however this is to be expected. The spin-polaron also lives on a
lattice with a two-site basis, so one expects to see two bands for it,
if both happen to fit below the continuum, as is the case here and for
larger $J_0$ values. This is also in agreement with the
strong-coupling limit which also predicts the two bands, see
Eq. (\ref{b5}). From there, we see that this higher energy
spin-polaron state can also be thought of as a propagating singlet,
where the singlet again involves the electron in a particular
wavefunction and the spin at its center. The wavefunction involves the
four sites surrounding the spin, but with different
$\vec{k}$-dependent phases (or symmetry) than for the low-energy
branch.

\begin{figure}[t]
\includegraphics[width=0.95\columnwidth]{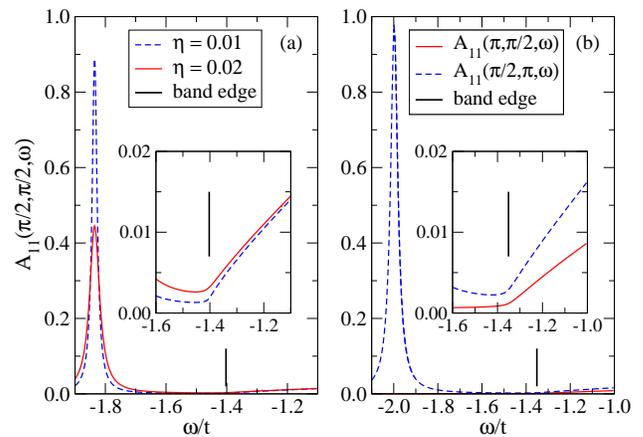}
\caption{(color online) (a) Spectral weight $A_{11}({\pi\over 2}, {\pi
  \over 2}, 
  \omega)$ as a function of energy, for $\eta = 0.01$ (dashed line)
  and $\eta = 0.02$ (full line). (b) $A_{11}({\pi}, {\pi \over 2},
  \omega)$ (full line) and $A_{11}({\pi\over 2}, {\pi }, \omega)$
  (dashed line) vs. $\omega$, for $\eta=0.02$. In both cases the
  vertical lines show the expected location of the lower band-edge of
  the continuum.  The insets show the same quantities, but
 with the focus on the lower band-edge of the continuum. The parameters
  are $t=1, J_0=5, 
  J=0.05$ and $S={1\over 2}$.}
\label{fig6}
\end{figure}

Besides these two spin-polaron bands, Fig. \ref{fig5} reveals another
dispersing feature at much higher energies, of order $-2t$. That this
is also a discrete (infinitely lived) state is demonstrated in
Fig. \ref{fig6}(a), where we verify its scaling with $\eta$: doubling
$\eta$ halves the height of the peak and doubles its width. Thus, this
is a Lorentzian of width $\eta$, and becomes a delta 
function as $\eta \rightarrow 0$. Note that
the continuum is independent of $\eta$ and begins at the expected
value, indicated by the vertical line (the inset focuses on this
feature and demonstrates this agreement more clearly). The continuum
has very low weight, 
explaining why it is mostly invisible on the scale of Fig. \ref{fig5}.

 Interestingly, we only see this higher-energy discrete state in some regions of the
BZ -- for example it is absent along the $(0,0)-(0,\pi)$ line (to be
more precise, we have searched with an $\eta$ as low as $10^{-4}$ and
did not see any features just below the gap. Of course, this does not
rule out a state with an extremely low $qp$ weight).  The apparent
``disappearance'' near $(\pi,\pi)$ is due to vanishing $qp$ weight,
but the state exists in that region.  Another interesting feature is
demonstrated by Fig. \ref{fig6}(b), which shows
$A_{11}(\vec{k},\omega)$ at two points that one would normally
consider equivalent, namely $\left(\pi, {\pi\over 2}\right)$ and
$\left( {\pi\over 2}, \pi\right)$. While the former shows a big $qp$
peak just above $-2t$, the later shows no weight at this energy. The
explanation is that by symmetry one needs to have
$A_{11}(k_x,k_y,\omega) = A_{22}(k_y,k_x,\omega)$, since ``1'' and
``2'' refer to states on the $x$, respectively $y$-rungs. However, for
$k_x\ne k_y$, there is no requirement that $A_{11}(k_x,k_y,\omega) =
A_{11}(k_y,k_x,\omega)$ and indeed we see that this is not the
case. Fig. \ref{fig6}(b) thus suggests that the disappearance of this
$qp$ state along some directions is due to symmetries, which result in
orthogonality between the spin-down free electron state and the true
eigenstate. Polarization-dependent spectroscopies should be able to
detect these variations between $A_{11}$ and $A_{22}$.

Regarding the origin of this high-energy $qp$ state, this seems to be
standard polaron physics. Holstein polarons at strong couplings are
also known to have a so-called ``second bound state''.\cite{MB1} It is
essentially an internal excited state of the polaron, possible if the
interaction and so the binding energy with its cloud is strong enough.

These two examples, together, are enough to give us intuition about
the evolution of the spectral weight in such models as $J_0$
increases. For small 
values only the lowest spin-polaron band is visible in regions where
it is well below the continuum. In the regions where the continuum
forces it to flatten, the $qp$ is extremely small. As $J_0$ increases,
more and more of the low-energy spin-polaron band emerges below the
continuum.  Then the upper spin-polaron band starts to emerge as well
and the part that is below the continuum is clearly visible, the rest
is flattened and with low $qp$ weight.  This is the case {\em e.g.}
for $J_0=3$ (not shown) where the entire lower band and about one
third of the upper band are visible below the continuum. For even
larger $J_0$ both spin-polaron bands are visible, and at yet higher
values, more bound states can split-off from the continuum.  The
number of $qp$ bands and their weights is therefore very sensitive to
strength of the coupling.

\begin{figure}[t]
\includegraphics[width=0.5\columnwidth]{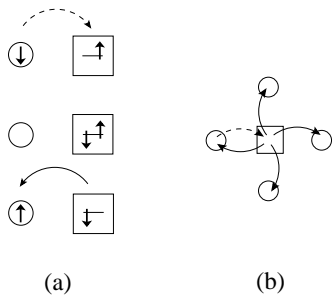}
\caption{(a) Illustration of exchange between the charge carrier and
  the impurity spin (depicted as a box), arising from virtual hoping of the
charge carrier to the spin site, and back. (b) The same type of
process can result in  charge carrier hopping with spin flip, if the
charge carrier moves to another site.}
\label{fig7}
\end{figure}

Before concluding, we must make one more important comment.
In the model considered here we assumed that the charge carrier
interacts with the spins only through exchange. This exchange comes
from virtual hopping of the charge carrier to the spin site and
back. Fig. \ref{fig7}(a) illustrates one such possible scenario,
leading to AFM exchange if a single half-filled orbital  gives
rise to the impurity spin. However, the charge carrier could hop off
to any of the nearest neighbor sites of the impurity spin, since it does not
necessarily have to return to the original site. Such processes are
illustrated in 
Fig. \ref{fig7}(b) and result in effective charge carrier hopping
with possible spin-flip correlated with a spin flip of the impurity
spin. The energy scale for such processes is the same as for the AFM
exchange, however their sign can be either negative or positive,
depending on whether the hopping integrals along the relevant links
have the same, or opposite signs. This, of course, depends on the
orbitals involved both at the charge carrier and the spin site.

Such hopping + spin-flip processes can be treated exactly in our
approach, as already stated. They can have dramatic effects, as we
illustrate now with a simple example. Consider the spin $S$ at site
$(ij)$ and its
four neighboring charge sites, which we label anticlockwise as $1= (ij,1)$,
$2=(ij,2)$, $3=(i-1,j,1)$ and $4=(i,j-1,2)$. Four symmetric linear
combinations can be built from these, namely
$\chi_{s\sigma}={1\over 2} (c_{1\sigma} + c_{2\sigma} + c_{3\sigma} +
c_{4\sigma})$; $\chi_{d\sigma}={1\over 2} (c_{1\sigma} - c_{2\sigma} +
c_{3\sigma} - c_{4\sigma})$; $\chi_{p1\sigma}={1\over 2} (c_{1\sigma}
+ c_{2\sigma} - c_{3\sigma} -c_{4\sigma})$ and
$\chi_{p2\sigma}={1\over 2} (c_{1\sigma}- c_{2\sigma} - c_{3\sigma}
+c_{4\sigma})$. Consider the impurity spin lowering part of the
Hamiltonian we have worked with so far, 
$
J_0 S^-(c^\dagger_{1\uparrow} c_{1\downarrow} + c^\dagger_{2\uparrow}
c_{2\downarrow}+ c^\dagger_{3\uparrow} c_{3\downarrow}+
c^\dagger_{4\uparrow} c_{4\downarrow}) = J_0 S^-
(\chi^\dagger_{s\uparrow} \chi_{s\downarrow} +\chi^\dagger_{d\uparrow}
\chi_{d\downarrow} + \chi^\dagger_{p1\uparrow} \chi_{p1\downarrow}
+\chi^\dagger_{p2\uparrow} \chi_{p2\downarrow}) 
$
{\em i.e.} all these symmetrized orbitals interact equally strongly
with the impurity, and could participate in the ``singlet''. Which happens
to be the GS is chosen by the signs of the hoppings and charge
carrier, as discussed. If hopping + spin flip processes are
included, the results are very different. Assume for example that all
hopping integrals between the 4  sites and the impurity site have
the same sign. In this case, $J_0 S^-c^\dagger_{1\uparrow}
c_{1\downarrow} \rightarrow J_0 S^-(c^\dagger_{1\uparrow}+c^\dagger_{2\uparrow}+
c^\dagger_{3\uparrow}+c^\dagger_{4\uparrow}) c_{1\downarrow}$, since
as illustrated in Fig. \ref{fig7}(b), the charge carrier removed from
site 1 can emerge at any of the spin's neighbor sites. The same is
true for all other 
terms, and adding them together, we obtain a total equal to $4 J_0 S^-
\chi^\dagger_{s\uparrow} \chi_{s\downarrow}$. In other words, in this
case only the
$s$-symmetry orbital can participate in the singlet (which state is
selected depends, obviously, on the relative signs for the hopping
between charge sites and the impurity site). As a result, we
expect to see a single spin-polaron state  at $\vec{k}=0$,
corresponding to a ``singlet'' involving this allowed orbital, and
the other spin-polaron states  discussed for the simple model  will
vanish from the spectrum.

\section{Summary and conclusions}

To summarize, we claim that essentially any problem that involves a
charge carrier interacting with a fully polarized FM background can be
solved exactly, irrespective of the complexity of the sublattices
involved or the range of the exchanges or other complications. Of
course, calculations become more complex as one makes the model more
complicated, but the exact solution following the approach suggested
here should always be possible.

We exemplified it on a relatively simple, yet interesting
two-dimensional case which we believe illustrates most of the physics
that can be expected in such systems. It allows us to elucidate the
nature of the spin-polaron quasiparticle, which at strong coupling is
a singlet (more generally, the maximally polarized state with spin
$S-{1\over 2}$) similar to the Zhang-Rice solution. However, we show
that the orbitals participating to the electronic wavefunction that
locks into the singlet with the lattice spin, have phases that vary as
a function of the singlet's momentum. In the ground state they have
the signs consistent with the symmetry of the orbitals involved and
also determined by the type of charge carrier. Moreover, since on
complex lattices one 
can form multiple such linear combinations which can participate in
the singlet, it is possible to see more than one spin-polaron band
below the continuum. If however hopping + spin-flip terms are
included, some of these spin-polaron bands will be removed because of
symmetry considerations.

We hope that these insights may help solve some of the known issues
regarding the normalization of the ZR singlet, which is an essential
ingredient in the cuprate physics. Extensions of the solution to AFM
backgrounds are very desirable, although clearly much more
complicated. We plan to investigate  such problems next.

As far as uses of doped FM insulators for spintronics applications are
concerned, this  work suggests that materials with an AFM exchange
$J_0$ between the charge carriers and the local spins could be very
interesting. For instance, consider a pulse of unpolarized 
charge carriers created  in such a medium, for instance by optical
means. If they are placed in an electric field, the charge carriers with 
spin parallel to the FM background will propagate with a different
speed than the charge carriers with spin antiparallel to the FM
background, since the latter are dressed by magnons and become
heavier spin-polarons while the former propagate as bare particles. 
Both types propagate coherently (at least at $T=0$) so this suggests 
that in time a pulse of unpolarized carriers will separate spatially
into two spin-polarized pulses traveling at different speeds.
Ingenious use of such difference between the two spin polarizations
may open the way towards using such materials as sources or detectors
of spin-polarized currents, which are essential components for
spintronic devices. Of course, issues such as the effect of finite-T
and finite charge carrier concentrations need to be understood first,
however this seems to be a promising line of investigation.

\section*{Acknowledgments}
This work was supported by NSERC  and the CIFAR Nanoelectronics (M.B.)
and Quantum Materials (G.S.) programs. Part of this work was carried
out at the Kavli Institute for Theoretical Physics, which is supported 
by  NSF under Grant No. PHY05-51164. 

\bibliographystyle{revtex}

\end{document}